\documentclass{INTERSPEECH2023}

\interspeechcameraready 

\usepackage{bm}
\usepackage{booktabs}
\usepackage{bbm}
\usepackage{setspace}
\usepackage{amsfonts,amssymb}
\usepackage{makecell}
\usepackage{multirow}
\usepackage{color}
\usepackage[capitalize]{cleveref}
\crefname{section}{Sec.}{Secs.}
\Crefname{section}{Section}{Sections}
\Crefname{table}{Table}{Tables}
\crefname{table}{Tab.}{Tabs.}

\title{Attention-based Interactive Disentangling Network for Instance-level Emotional Voice Conversion}
\name{Yun Chen$^{1}$, Lingxiao Yang$^{1}$, Qi Chen$^{1}$, Jian-Huang Lai$^{1,2,3}$ and Xiaohua Xie$^{1,2,3*}$\thanks{*Corresponding Author}}
\address{\textsuperscript{1}School of Computer Science and Engineering, Sun Yat-Sen University, China \\
\textsuperscript{2}Guangdong Province Key Laboratory of Information Security Technology, China \\
\textsuperscript{3}Key Laboratory of Machine Intelligence and Advanced Computing, Ministry of Education, China}
\email{\{cheny2259, chenq377\}@mail2.sysu.edu.cn, \{yanglx9, stsljh, xiexiaoh6\}@mail.sysu.edu.cn}

\begin{document}

\maketitle
 
\begin{abstract}

Emotional Voice Conversion aims to manipulate a speech according to a given emotion while preserving non-emotion components.
Existing approaches cannot well express fine-grained emotional attributes.
In this paper, we propose an \textit{\textbf{A}ttention-based \textbf{I}nteractive dise\textbf{N}tangling \textbf{N}etwork (AINN)} that leverages instance-wise emotional knowledge for voice conversion.
We introduce a two-stage pipeline to effectively train our network: Stage I utilizes inter-speech contrastive learning to model fine-grained emotion and intra-speech disentanglement learning to better separate emotion and content.
In Stage II, we propose to regularize the conversion with a multi-view consistency mechanism. 
This technique helps us transfer fine-grained emotion and maintain speech content.
Extensive experiments show that our AINN outperforms state-of-the-arts in both objective and subjective metrics. 

\end{abstract}

\vspace{-6pt}
\section{Introduction}

\begin{figure}[t]
    \centering
    \includegraphics[width=0.44\textwidth]{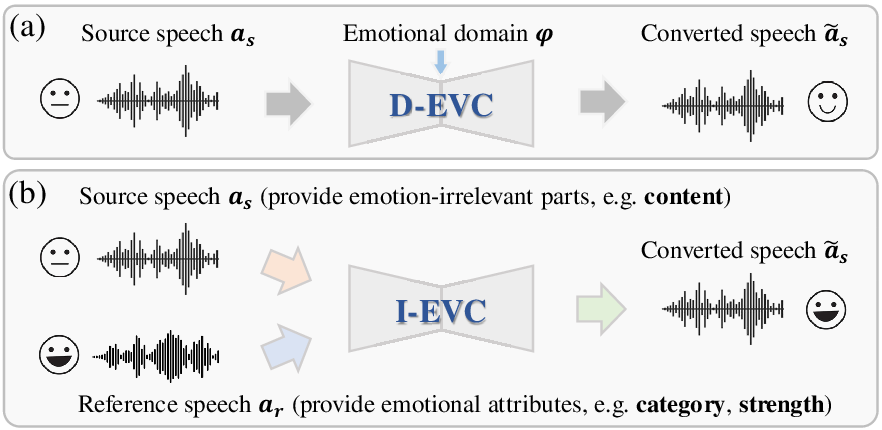}
    \vspace{-5pt}
    \caption{Illustration of (a) Domain-level emotional voice conversion, and (b) Instance-level emotional voice conversion.}
    \label{fig:fig1}
    \vspace{-15pt}
\end{figure}

Emotional Voice Conversion (EVC) aims to transfer the emotion of a speech while preserving emotion-independent components.
EVC has garnered much attention due to its potential applications in areas such as virtual assistants \cite{elgaar2020multi}, voice cloning \cite{chen2022v2c}, and human-computer interaction \cite{zhou2021review}.
The recent advancement in EVC is driven by the use of Generative Adversarial Networks (GANs) \cite{zhou2020transforming, rizos2020stargan, shankar2020non}, Autoencoder \cite{gao2019nonparallel, zhou2020converting, cao2020nonparallel, zhou2021vaw, zhou2021seen}, and Sequence-to-sequence model \cite{robinson2019sequence, kim2020emotional, zhou21b_interspeech}.
In general, most of existing EVC methods take an emotional domain descriptor $\bm{\phi}$ (emotion label or a pre-defined emotion embedding) as a condition to guide the input speech $\bm{a}_{s}$ for emotion conversion (\cref{fig:fig1}(a)).
However, since emotions have varying degrees in speech, it is difficult to capture fine-grained emotions using only the domain descriptor.
We define this task as \textbf{Domain-level emotional voice conversion (D-EVC)}: $\tilde{\bm{a}}_{s} = \mathcal{M}(\bm{a}_{s}, \bm{\phi})$.

To achieve the fine-grained expression, we propose to extract instance-level emotional representation (\textit{e.g.} emotional category, strength) from a reference speech $\bm{a}_{r}$, and then incorporate it with the emotion-independent components of source speech $\bm{a}_{s}$: $\tilde{\bm{a}}_{s} = \mathcal{M}(\bm{a}_{s}, \bm{a}_{r})$.
We  refer to this task as \textbf{Instance-level emotional voice conversion (I-EVC)} (\cref{fig:fig1}(b)).
Due to the representation gap between the emotional domain and speech instance, I-EVC faces two challenges:
1) Unlike D-EVC which provides pure emotional category, I-EVC needs to capture the emotion-related information from a reference speech and avoid the interference of emotion-irrelevant components.
2) Unlike the well-defined emotion category, emotional strength is continuous and highly 
subjective, making it hard to be modeled without supervised signals.
StarGANv2-VC \cite{li2021starganv2} achieves instance-level conversion by involving a reference speech for style representation.
Nevertheless, this approach lacks strength-related modeling, resulting in uncontrolled strength variations in the converted speech.
Recently, Emovox \cite{zhou2022emotion} extracts strength value by the relative function \cite{parikh2011relative} and uses it to help strength transfer. 
However, it does not explicitly establish the relationship between pre-calculated strength and high-level emotional speech features, limiting the strength conversion.
Besides, Emovox requires additional text transcriptions to disentangle the emotional feature, which is less flexible for low-resourced languages \cite{lee2021voicemixer}.

\begin{figure*}[t]
    \centering
    \includegraphics[width=\textwidth]{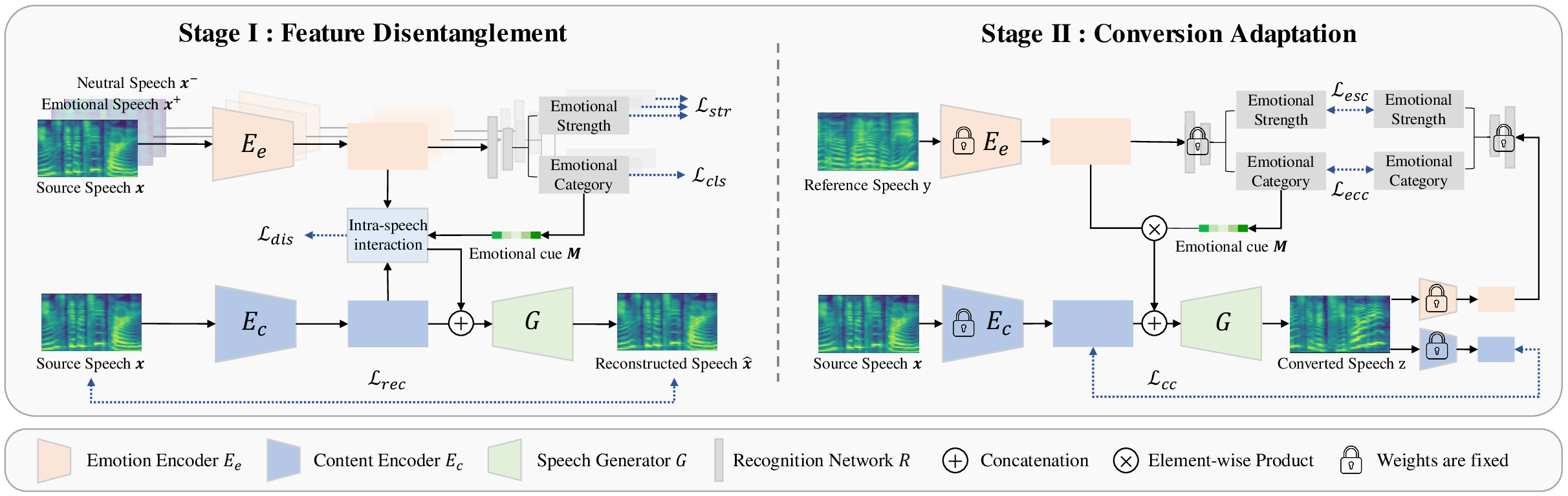}
    \vspace{-12pt}
    \caption{The framework of the proposed method. It consists of two stages: Feature disentanglement and Conversion adaptation.}
    \label{fig:fig2}
    \vspace{-5pt}
\end{figure*}

To address instance-level emotional voice conversion, we propose an \textit{\textbf{A}ttention-based \textbf{I}nteractive dise\textbf{N}tangling \textbf{N}etwork (AINN)} with a two-stage training pipeline.
The first stage aims to disentangle emotion and emotion-irrelevant features under the auto-encoder paradigm.
Specifically, we contrastively learn the relation of inter-speech to model emotional strength and construct complete emotion representation.
Meanwhile, we establish an attention-based method to effectively decompose the emotion and content of a speech.
In the second stage, multi-view consistency is applied to enable the model to learn more task-adaptive knowledge. 
Our method ensures that the speech content remains nearly unchanged, while the fine-grained emotion is improved.
The main contributions of this paper include: 
\begin{itemize}
    \item We propose an Attention-based Interactive diseNtangling Network (AINN) with an efficient two-stage training strategy for instance-level emotional voice conversion. 
    \item We propose a text-free feature disentanglement method that utilizes attention-based emotional cues to better decompose emotion and content.
    \item We incorporate contrastive learning to effectively model emotional strength, and leverage it to improve speech consistency on fine-grained emotion.
    \item Extensive experiments demonstrate the effectiveness of our method in both objective and subjective metrics. 
\end{itemize}

\section{Method}
Given a source speech, our goal is to generate a new speech, where we can adjust the emotional category and strength of it towards a reference speech, and ensure its content remains unchanged. 
We expect the emotional attributes to be only guided by reference speech and the content aligns with the source speech.
To this end, we design a two-stage training pipeline: \textit{Feature disentanglement} and \textit{Conversion adaptation}. 
The overview of our framework is shown in \cref{fig:fig2}.
In Stage I, We train the whole network to disentangle the emotional and content features from the input speech mel-spectrogram in an auto-encoder manner.
Specifically, we introduce an \textit{inter-speech interaction} method to model instance-level emotion and an \textit{intra-speech interaction} method to enforce independent content and calibrated emotion.
Stage II only fine-tunes the generator to fuse two parts of features and synthesize reasonable speech in the conversion progress.
It is achieved by \textit{multi-view consistency} including emotional category, strength, and content.

\subsection{Stage I: Feature disentanglement}
\label{sec:2.1}

\subsubsection{Inter-speech interactive emotion representation}
\label{sec:2.1.1}
Given a source speech mel-spectrogram $\bm{x}$, we can obtain an emotional feature $\bm{e} = E_e(\bm{x})$.
To make the emotion encoder $E_e$ capture rich fine-grained emotional features, we feed $\bm{e}$ into the two-branch recognition network $R=\{R_{c}, R_{s}\}$ and force it to achieve \textit{emotion classification} and \textit{strength assessment}.

\noindent\textbf{Emotional classification.} 
For $K$ emotional category, we adopt widely used cross-entropy loss, which is defined as follows:
\begin{equation}
     \mathcal{L}_{cls} = \mathbb{E}_{\bm{x}}[- \sum_{k=1}^K \bm{p}_{k} \ log(\bm{q}_{k})],
     \label{eq:eq1}
\end{equation}
where $\bm{p}$ represents the one-hot ground truth and ${\bm{q}} = R_{c}(\bm{e})$ indicates the classification probability distribution.

\noindent\textbf{Strength assessment.}
Due to lack of annotations, it is difficult to directly apply supervision to emotional strength.
Therefore, we adopt a self-supervised approach to assess strength by mining relative relations \cite{parikh2011relative} between speeches. 
In previous works \cite{zhou2022emotion, zhu2019controlling, li2022cross}, neutral speech is often treated as non-emotion speech, and emotional strength is considered as relative difference between neutral speech and emotional speech.
Based on that prior, we build relative relations with two hypotheses:
1) The strength of neutral speech approaches zero, and the strength of emotional speech is significantly higher than that of neutral speech.
2) The strength difference between emotional speeches is small.
Therefore, we model the relative relation as a triplet $\{\bm{x}, \bm{x}^{+}, \bm{x}^{-} \}$, where the positive pair $\{\bm{x}, \bm{x}^{+}\}$ shares the same emotional category, and the negative pair $\{\bm{x}, \bm{x}^{-}\}$ contains an emotional speech mel-spectrogram $\bm{x}$ and a neutral speech mel-spectrogram $\bm{x}^{-}$.
We set the strength of $\bm{x}^{-}$ to zero, then the triplet contrastive loss is applied to narrow the strength difference of the positive pair, and enlarge that of the negative pair:
\begin{equation}
    \begin{split}
        \mathcal{L}_{str} = & \ \mathbb{E}_{x, x^{+}, x^{-}}[\ max \{|s^{x} - s^{x^{+}}|,\ \delta_1 \} \\ & + \ max \{ 1 - (s^{x} - s^{x^{-}}),\ \delta_2 \}],
    \end{split}
     \label{eq:eq2}
    \end{equation}
where $s=R_{s}(\bm{e})$ is emotional strength, and $s^{x}$, $s^{x^{+}}$, $s^{x^{-}}$ are the strengths of $\bm{x}$, $\bm{x}^{+}$, $\bm{x}^{-}$, respectively. 
Two margins $\delta_1 $ and $\delta_2$ are employed to prevent the network from overfitting to extreme strength. 
The incorporation of strength learning allows the network to acquire knowledge related to strength, enhancing personalized information in emotional feature.
This learned network is also used to predict reasonable strength values for consistency learning in Stage II (\cref{sec:2.2}).

\subsubsection{Intra-speech interactive disentanglement}
\label{sec:2.1.2}
\begin{figure}[t]
    \centering
    \includegraphics[width=0.42\textwidth]{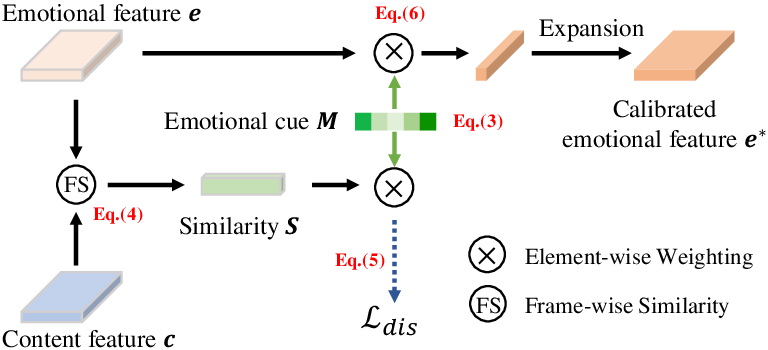}
    \caption{Illustration of Intra-speech interactive method.}
    \label{fig:fig3}
    \vspace{-15pt}
\end{figure}

To achieve effective disentanglement over emotional and content features, it is crucial to identify meaningful frames in a speech.
These frames should have two attributes: 
1) They should not be silent frames as they do not contain useful information. 
2) The emotion expressed in these frames should be clearly perceivable so that the network can focus on the significant difference between emotional and content features.
In this section, we first introduce an attention mechanism that helps locate these meaningful frames in the speech.
And then, the feature disentanglement can be realized by minimizing the similarity between emotional feature $\bm{e}=E_e(\bm{x})$ and content feature $\bm{c}=E_c(\bm{x})$.
This process is shown in \cref{fig:fig3}.

Firstly, inspired by the idea of class activation mapping (CAM) \cite{zhou2016cvpr,Chen2022sipe}, we estimate an emotional cue $\bm{M}$ to indicate the meaningful frames:
\begin{equation}
    \bm{M} = ReLU(\bm{\theta}^T \cdot \bm{e}),
     \label{eq:eq3}
\end{equation}
where $\bm{\theta}$ is the classifier weight of $R_{c}$, relating to the corresponding emotional category.
$\bm{M}$ is further normalized to [0, 1] by the maximum value along the temporal axes.
Since $\bm{M}$ is class-aware attention, it naturally filters silent frames that are not related to emotion, and concentrates on meaningful frames.

Secondly, we calculate the frame-wise similarity $\bm{S}$ for emotional feature $\bm{e}$ and content feature $\bm{c}$:
\begin{equation}
    \bm{S}_t = \frac{|e_{t} \cdot c_t|}{||e_t||_2 \cdot ||c_t||_2}, 
     \label{eq:eq4}
\end{equation}
where $t \in [0, T] $ indexes the $t$-th frame of the feature.
Thus the feature disentanglement loss is defined as:
\begin{equation}
    \mathcal{L}_{dis} = \mathbb{E}_{\bm{x}} \ [\frac{1}{T} \sum_{t=0}^T{\bm{M}_t \cdot \bm{S}_t}\ ].
     \label{eq:eq5}
\end{equation}

Since emotional cues give high responses on frames with the discriminative emotional tone, we calibrate the emotional feature by emotional cues:
\begin{equation}
    \bm{e}^{*} = \frac{1}{T}\sum_{t=0}^T{\bm{M}_t \cdot e_t}.
     \label{eq:eq6}
\end{equation}

To train the network, we employ the auto-encoder manner to reconstruct the speech mel-spectrogram $\bm{x}$ by a generator $G$ and define a reconstruction loss as:
\begin{equation}
    \mathcal{L}_{rec} = \mathbb{E}_{\bm{x}}[||G(\bm{c}, \bm{e}^{*}) - \bm{x}||_1].
     \label{eq:eq7}
\end{equation}

\noindent\textbf{Full objectives.} The total loss of Stage I is as follows:
\begin{equation}
    \mathcal{L}_{stage1} = \mathcal{L}_{rec} + \mathcal{L}_{cls} + \lambda_{dis} \mathcal{L}_{dis} + \lambda_{str} \mathcal{L}_{str},
     \label{eq:eq8}
\end{equation}
where $\lambda_{dis}$ and $\lambda_{str}$ are hyperparameters for balancing losses.

\subsection{Stage II: Conversion adaptation}
\label{sec:2.2}
Given the fixed emotion encoder $E_{e}$, content encoder $E_{c}$, and recognition network $R$ pre-trained in Stage I, our second stage is to fine-tune only the generator $G$ to adapt to conversion using a style transfer paradigm \cite{johnson2016perceptual,zhang2022exploring}.
Owing to the pre-trained networks that provide a good feature disentanglement for speech, 
our framework incorporates a multi-view consistency using well-disentangled features, improving the expression of fine-grained emotions and preserving the integrity of speech content.
As shown in the right part of \cref{fig:fig2}, our network takes a source speech mel-spectrogram $\bm{x}$ and a reference speech mel-spectrogram $\bm{y}$ as input.
The network extracts their content feature $E_{c}(\bm{x})$ and the calibrated emotional feature $E_e(\bm{y})^{*}$.
Then the generator $G$ produces the converted mel-spectrogram $\bm{z} = G(E_{c}(\bm{x}), {E_{e}(\bm{y})}^{*})$.
We constrain $\bm{z}$ to be similar to $\bm{x}$ and $\bm{y}$ in terms of content and emotion, respectively.

\noindent\textbf{Content consistency.}
The converted speech mel-spectrogram $\bm{z}$ is required to preserve the content feature of $\bm{x}$.
Following the commonly used constraint in voice conversion \cite{qian2019autovc}, we employ a consistency loss $\mathcal{L}_{cc}$ to prevent the content from being lost:
\begin{equation}
    \mathcal{L}_{cc} = \mathbb{E}_{\bm{x}, \bm{y}}[||E_c(\bm{z}) - E_c(\bm{x})||_1].
     \label{eq:eq9}
\end{equation}
\noindent\textbf{Emotion consistency.}
Moreover, compared to content information, the emotional feature tends to be more global in a speech.
Therefore, we apply two consistency losses to regulate the emotional attributes from a global perspective.

First, we choose KL-divergence as emotional category consistency loss $\mathcal{L}_{ecc}$ to align two distributions:
\begin{equation}
    \mathcal{L}_{ecc} = \mathbb{E}_{\bm{x}, \bm{y}} [ -\sum_{}{R_c(E_{e}(\bm{y})) \ log\frac{R_c(E_{e}(\bm{z}))}{R_c(E_{e}(\bm{y}))}}].
     \label{eq:eq10}
\end{equation}
In addition, an MSE-based emotional strength consistency loss $\mathcal{L}_{esc}$ is adopted to regress the assessed value, shown as:
\begin{equation}
    \mathcal{L}_{esc} = \mathbb{E}_{\bm{x}, \bm{y}} {[||R_s(E_{e}(\bm{z})) - R_s(E_{e}(\bm{y}))||^2_{2}]},
     \label{eq:eq11}
\end{equation}
\noindent\textbf{Full objectives.} The total loss of Stage II is defined as:
\begin{equation}
     \mathcal{L}_{stage2} = \mathcal{L}_{rec} + \lambda_{c} \mathcal{L}_{cc} + \lambda_{c} \mathcal{L}_{ecc} + \lambda_{c} \mathcal{L}_{esc}.
     \label{eq:eq12}
\end{equation}
where $\mathcal{L}_{rec}$ is the reconstruction loss to ensure the stability of converted speech.
$\lambda_{c}$ is a hyperparameter for balancing loss.

\setlength{\tabcolsep}{3.5mm}
\begin{table*}[!t]
    \centering
    \caption{Quantitative comparisons of converted speech with previous methods. The best results are shown in bold. The * denotes methods with text supervision and pretrained on VCTK dataset \cite{veaux2017cstr}.}
    \vspace{-6pt}
    \begin{tabular}{c|c|ccc|cc}
    \Xhline{1pt}
    \multirow{2}{*}{Methods} & \multirow{2}{*}{Type} & \multicolumn{3}{c|}{Objectiveness} & \multicolumn{2}{c}{Subjectiveness} \\
    \cline{3-7}    & & MCD $\downarrow$ & $\rm{ACC}_{{cls}}$ $\uparrow$ & $\mathrm{RMSE_{{str}}}$ $\downarrow$ & Naturalness $\uparrow$ & Similarity $\uparrow$\\
    \hline
    EmotionalStarGAN \cite{rizos2020stargan} & Domain-level & 5.219 & 0.743 & 0.312 & 2.78 $\pm$ 0.56 & 48.37\% \\
    Seq2Seq-EVC* \cite{zhou21b_interspeech} & Domain-level & 5.323& 0.550 & 0.215 & 2.62 $\pm$ 0.54 & 55.82\%\\
    \hline
    Emovox* \cite{zhou2022emotion} & Instance-level & 5.088& 0.460 & 0.199 & 2.96 $\pm$ 0.53 & 47.67\%\\
    AINN (Ours) & Instance-level & \textbf{4.596}& \textbf{0.830} & \textbf{0.117} & \textbf{3.43 $\pm$ 0.59} & \textbf{76.12}\%\\
    \Xhline{1pt}
    \end{tabular}
    \label{tab:tab1}
    \vspace{-11pt}
\end{table*}

\vspace{-6pt}
\section{Experiments}
\subsection{Experimental Setup}
\textbf{Dataset.} We train our model on 10 English speakers from the Emotional Speech Dataset (ESD) \cite{zhou2021seen} with five emotions.
Each speaker includes 300 training, 20 validation, and 30 test speeches. 
In our experiments, all speech data is utilized in the form of an 80-dimensional mel-spectrogram extracted from the audio with a sampling rate of 16kHz. 
We also use a pre-trained HiFi-GAN \cite{kong2020hifi} to synthesize audio from mel-spectrogram.

\noindent\textbf{Data preparation.} To adapt to the I-EVC task, we construct source-reference speech pairs as input.
The source and reference speeches are sampled from different emotional categories of the same speaker.
We create 20 conversion types between five emotions, such as neutral to happy and sad to angry.
For each conversion type, we randomly sample 750 speech pairs for training, 50 pairs for validation, and 15 pairs for testing.

\noindent\textbf{Network architecture.} The architecture of the content encoder and generator are based on AutoVC \cite{qian2019autovc}, and the emotion encoder consists of six convolution layers and an LSTM layer.
The recognition network is a $1 \times 1$ convolution layer.
More details are given in the supplementary material.

\noindent\textbf{Implementation details.}
We adopt Adam optimizer with a learning rate of 1e-4 for Stage I and 1e-5 for Stage II.
The network is trained with a batch size of 64 for 50000 iterations in both stages.
The total training time is about nine hours on an NVIDIA A100 GPU.
For the loss functions in \cref{eq:eq8} and \cref{eq:eq12}, we set $\lambda_{dis}=0.2$, $\lambda_{str}=1$, $\lambda_{c}=0.0002$.
To avoid strength collapsing, those two margins in \cref{eq:eq2} are set to 0.5.
Our code and converted samples are available at \url{https://ainn-evc.github.io/}.

\setlength{\tabcolsep}{1.8mm}
\begin{table}[t]
    \centering
    \vspace{-6pt}
    \caption{Ablation study of the methods.}
    \vspace{-6pt}
    \begin{tabular}{l|ccc}
    \Xhline{1pt}
    Methods & MCD $\downarrow$ & $\mathrm{ACC_{{cls}}}$ $\uparrow$ & $\mathrm{RMSE_{{str}}}$ $\downarrow$\\
    \hline
    baseline ($\mathcal{L}_{rec}$ + $\mathcal{L}_{cls}$) & 4.922 & 0.027 & 0.446\\
    + Stage I (w/o EC) & 4.766 & 0.323 & 0.283\\
    + Stage I (w/ EC) & \textbf{4.493} & 0.647 & 0.146\\
    \hline
    + Stage I \& II (\textbf{Full}) & 4.596 & \textbf{0.830} & \textbf{0.117}\\
    
    \Xhline{1pt}
    \end{tabular}
    \label{tab:tab2}
    \vspace{-16pt}
\end{table}

\vspace{-6pt}
\subsection{Comparison with previous methods}

Our method is compared with three state-of-the-arts methods, including domain-level EmotionalStarGAN \cite{rizos2020stargan} and Seq2Seq-EVC \cite{zhou21b_interspeech}, and instance-level Emovox \cite{zhou2022emotion}.

\noindent\textbf{Objective Evaluation.}
Following prior works \cite{rizos2020stargan, zhou2021seen}, we adopt Mel-cepstral distortion (MCD) \cite{kubichek1993mel} as a basic evaluation metric for evaluating content preservation.
Besides, we use a pre-trained emotion classifier ACRNN \cite{chen20183} to evaluate the accuracy (ACC$\mathrm{_{cls}}$) of converted speech. 
Furthermore, we evaluate the RMSE of strength between reference speech and converted speech.
The strength value is obtained by our strength assessment network $R_{s}$.
Note that the strength assessment network is frozen in Stage II, it would not be impacted by our converted speech, thereby a fair comparison can be drawn.
As shown in \cref{tab:tab1}, the proposed AINN achieves the best score in MCD, which indicates that our model effectively transforms emotion while maintaining the content.
Similarly, our AINN improves the classification accuracy by up to 83\%, outperforming all compared methods.
For RMSE$\mathrm{_{str}}$, Emovox obtains the best value 0.199 among all compared methods.
Our method, however, largely outperforms Emovox due to the proposed instance-level emotion modeling and multi-view consistency.

\noindent\textbf{Subjective Evaluation.}
We further conduct a user study to evaluate the naturalness and emotional similarity of converted speech.
Following previous works \cite{zhou2021seen, lee2021voicemixer, im2022emoq}, we recruit 15 subjects from the social network and they are proficient in English.
Each of them is asked to listen to 70 converted speeches.
From \cref{tab:tab1}, we note that the proposed method receives notable preference for both emotion similarity and overall performance.

\vspace{-6pt}
\subsection{Ablation Study}

We conduct ablation experiments to verify the contribution of each component in our method.
As shown in \cref{tab:tab2}, we define the model trained using only reconstruction and classification losses as the baseline method.
Compared to the baseline, the introduction of the feature disentanglement in Stage I lead to significant improvements across all evaluation metrics.
Furthermore, the use of emotional cues (EC) largely boosts the model's performance, with the ACC$\mathrm{_{cls}}$ increasing from 0.323 to 0.647 and RMSE$\mathrm{_{str}}$ decreasing to 0.146.
(The emotional cues are visualized in the supplementary material.)
This indicates that the attention mechanism successfully captures emotion-related information, resulting in better emotion representation.
Our full model with two stages increases the classification accuracy and narrows the RMSE of strength a lot with the contribution of multi-view consistency.
These results demonstrate that the two-stage training strategy with the well-designed disentanglement method and multi-view consistency is highly effective.

\vspace{-6pt}
\subsection{Feature Disentanglement}
Following prior work \cite{lee2021voicemixer}, we conduct emotion classification on the content feature of source speech to demonstrate the effect of feature disentanglement.
As shown in \cref{tab:tab3}, we observe a decline in the classification accuracy of content feature from 56.3\% to 34.3\% with the incorporation of the proposed methods. 
This proves the content feature does not contain emotion-related information.
By contrast, the classification accuracy of emotional feature remains consistently high, confirming that emotional information is preserved in feature disentanglement.

\setlength{\tabcolsep}{1.5mm}
\begin{table}[t]
    \centering
    
    \vspace{-6pt}
    \caption{Emotion classification accuracy on content feature and emotional feature of source speech.}
    \vspace{-6pt}
    \begin{tabular}{l|cc}
    \Xhline{1pt}
    Methods & Content feature $\downarrow$ & Emotional feature $\uparrow$\\
    \hline
    baseline & 0.563 & 0.903\\
    Stage I (w/o EC) & 0.395 & 0.904\\
    Stage I (\textbf{Full}) & \textbf{0.343} & \textbf{0.915}\\  
    \hline
    
    \Xhline{1pt}
    \end{tabular}
    \label{tab:tab3}
\end{table}

\begin{figure}[t]
    \centering
    \vspace{-5pt}
    \includegraphics[width=0.47\textwidth]{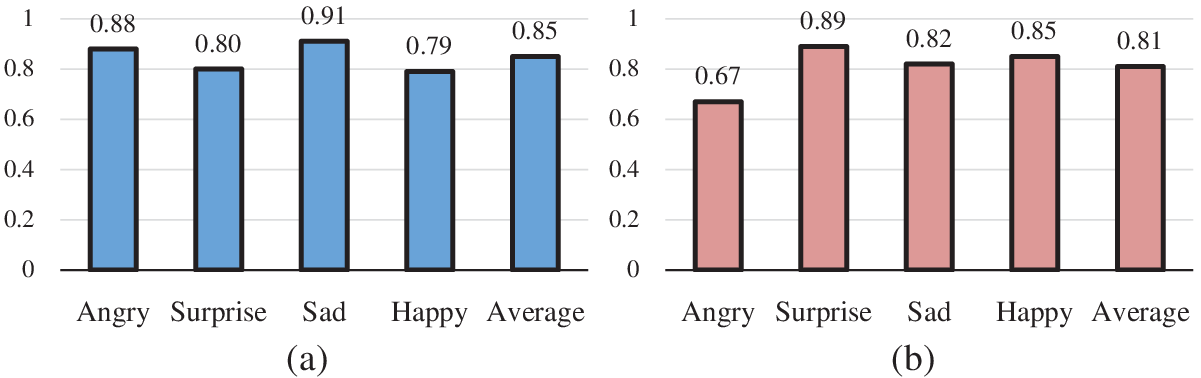}
    \vspace{-20pt}
    \caption{The strength accuracy of (a) source speech and (b) converted speech.}
    \label{fig:fig4}
    \vspace{-18pt}
\end{figure}

\vspace{-6pt}
\subsection{Strength Learning}
We perform a subjective experiment to investigate the effectiveness of the proposed strength learning method.
We create two groups of speech data, one including 20 pairs of real speeches and the other including 20 pairs of converted speeches.
Each pair contains two emotional speeches sharing the same emotion with varying emotional strengths.
Subjects are asked to identify the more expressive speech in each pair.
The results show that the strength accuracy achieves 85\% in the real speech (\cref{fig:fig4} (a)) and 81\% in the converted speech (\cref{fig:fig4} (b)). 
This consistently high performance provides strong evidence for the correctness of our strength assessment.
The pitch contour of converted speech is shown in the supplementary material. 
Furthermore, we conduct an objective comparison between our full model and the model trained without strength learning.
In this case, the model without strength learning degrades to domain-level EVC, resulting in a slight increase (+0.027) in RMSE$\mathrm{_{str}}$.
In summary, the effectiveness of strength learning is demonstrated from both subjective and objective aspects. 

\section{Conclusions}
We propose an \textit{Attention-based Interactive diseNtangling Network (AINN)} and a two-stage training pipeline for instance-level emotional voice conversion.
Our method effectively models emotional strength, and disentangles emotional and content features with the help of emotional cues.
Moreover, we introduce multi-view consistency to enhance fine-grained emotions and preserve the integrity of speech content.
Experimental results achieve state-of-the-art in all metrics.
Future work includes improving the quality of converted speech and achieving cross-speaker instance-level emotional voice conversion.

\noindent \textbf{Acknowledgements.} This project is supported by Guangdong-Hong Kong-Macao Greater Bay Area International Science and Technology Innovation Cooperation Project (No.2021A0505030080).

\clearpage

\setcounter{page}{1}
{
   \newpage
       \twocolumn[
        \centering
        \Large
        \vspace{0.5em}\textbf{Supplementary Material} \\
        \vspace{1.0em}
       ] 
   }

\section{Model Architecture}

\begin{figure}[t]
    \centering
    \includegraphics[width=0.48\textwidth]{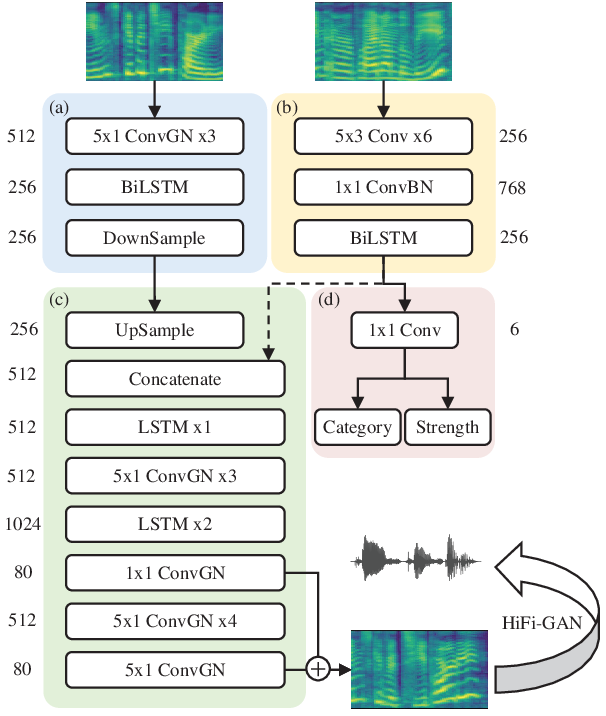}
    \caption{The architecture of our AINN.}
    \label{fig:arch}
\end{figure}

As shown in \cref{fig:arch}, our AINN consists of four networks: 
(a) a content encoder for extracting emotion-irrelevant features from the speech,
(b) an emotion encoder for extracting emotional attributes including categories and strengths,
(c) a generator for fusing feature embeddings and synthesizing reasonable speech,
and (d) a recognition network for classifying emotional categories and predicting the strength of emotional speech.
We re-calibrate the emotion embedding by emotional cue and then feed it to the generator.
For simplification, we indicate this process using a dotted line.
Finally, the HiFi-GAN synthesizes audio from the converted mel-spectrogram.
The total parameters of our network are 39.83 M.
ConvGN and ConvBN denote convolution followed by group normalization and batch normalization, respectively. 
BLSTM denotes bi-directional LSTM.
The number above each block represents the cell/output dimension of the structure. 

\section{Visualization}

\begin{figure}[t]
    \centering
    \includegraphics[width=0.48\textwidth]{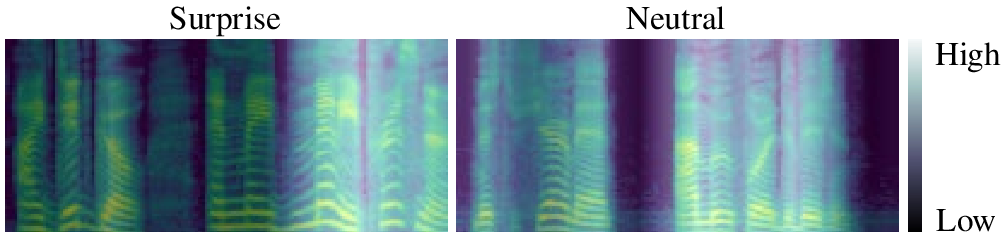}
    \caption{Visualization of emotional cue over spectrograms.}
    \label{fig:overlap}
\end{figure}

\noindent\textbf{Emotional Cue.}
We visualize the emotional cue over spectrograms to show its effectiveness.
As shown in \cref{fig:overlap}, we can see the emotional cue focuses on the significant ending of surprise speeches.
Also, it ignores the silent frames in the speaking gap, empowering a better emotional representation.

\begin{figure}[t]
    \centering
    \includegraphics[width=0.48\textwidth]{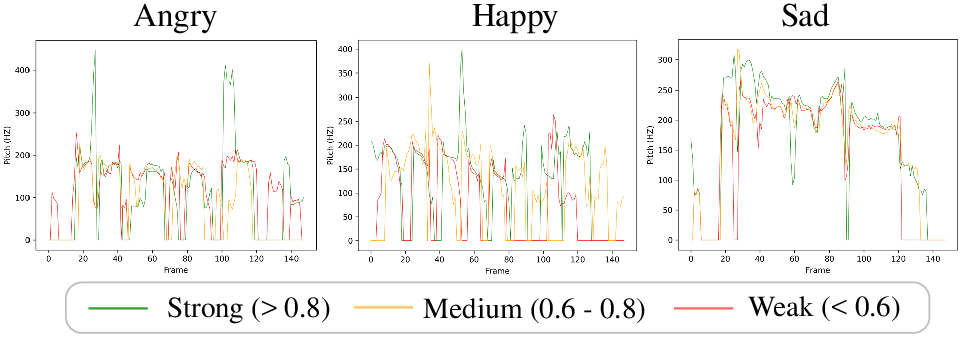}
    \caption{Pitch contours of converted speech with the same source speech for three emotions and three strength levels.}
    \label{fig:pitch}
\end{figure}

\noindent\textbf{Emotional Strength.}
We draw the pitch contour of converted speeches following \cite{zhu2019controlling, zhou2022emotion, li2022cross}.
As shown in \cref{fig:pitch}, we compare the pitch values of three converted speeches with varying strengths.
The comparison is performed on three groups of emotions. 
It can be observed that the converted speech with a larger strength score tends to exhibit a higher pitch value and larger fluctuation. 
Overall, this indicates the effectiveness of strength learning, as we are able to control the strength of the converted speech successfully.

\bibliographystyle{IEEEtran}
\bibliography{mybib}

\ifinterspeecharxiv \clearpage \input{appendix} \fi

\end{document}